\documentclass[12pt]{amsart}
\usepackage{latexsym}
\usepackage{amsfonts}
\usepackage{hyperref}
\usepackage{xcolor}

\usepackage[top=1in, bottom=1.2in, left=1in, right=1in]{geometry}

\newcommand{\Z}{\mathbb{Z}}

\newtheorem{remark}{Remark}

\author[]{Jiale Chen}
\address{Department of Mathematics, The City College of New York, New York,
NY 10031} \email{jchen056@citymail.cuny.edu}

\author[]{Dima Grigoriev}
\address{CNRS, Math\'ematiques, Universit\'e de Lille, 59655, Villeneuve d'Ascq, France}
\email{Dmitry.Grigoryev@univ-lille.fr}

\author[]{Vladimir Shpilrain}
\address{Department of Mathematics, The City College of New York, New York,
NY 10031}
\email{shpilrain@yahoo.com}

\begin{document}

\title[Digital signature schemes]{Digital signature schemes using non-square matrices or scrap automorphisms}

\begin{abstract}
We offer two very transparent digital signature schemes: one using non-square matrices and the other using scrap automorphisms. The former can be easily converted to a public key encryption scheme.
\end{abstract}

\maketitle

\section{Introduction}

Due to the concern that if large-scale quantum computers are ever built, they will compromise the security of many commonly used cryptographic algorithms, NIST had begun in 2016 a process to develop new cryptography standards and, in particular, solicited proposals for new digital signature schemes \cite{NIST} resistant to attacks by known quantum algorithms, such as e.g. \cite{Shor}.

One possible way to avoid quantum attacks based on solving the hidden subgroup problem (including the attacks in \cite{Shor}) is {\it not} to use one-way functions on sets that have an obvious (semi)group structure.
In this paper, we offer two digital signature schemes that are similar in spirit.
What they have in common is the following idea. There is a function $F: X \to Y$ that is not necessarily one-way. However, when only a {\it part} (call it $\bar{Y}$) of $Y$ is published, the restriction
$\bar{F}: \bar{X} \to \bar{Y}$ becomes much harder to invert and therefore becomes a candidate for a one-way function.

One of the simplest instantiations of this general idea is employed in our first digital signature  scheme. There, $X$ is an invertible square matrix over a commutative ring and $Y$ is $X^{-1}$. The function $F$ therefore takes a square matrix as the input and outputs the inverse matrix. This is often not a one-way function. However, if we only publish a subset of columns of $X^{-1}$, then finding the corresponding subset of rows of $X$ appears to be a hard problem. The corresponding digital signature scheme is described in Section \ref{description}. We note that the verification part in this scheme is very efficient since it amounts to multiplying a vector of sparse polynomials by a matrix whose entries are sparse polynomials (over $\Z_q$, for a small $q$).

For our second scheme, we consider an automorphism $\alpha: K \to K$ of an algebra $K$ of polynomials in $n$ variables $x_1, \ldots, x_n$ over a ring $R$. Such an automorphism can be given by $n$ polynomials $\alpha(x_i)$. It is arguable whether or not computing $\alpha^{-1}$ from all $\alpha(x_i)$ is a hard problem (this might depend on the ring $R$). Some of the known methods use Gr\"{o}bner bases techniques (see e.g. \cite{ShYu}) that are not known to be efficient. A more efficient algorithm was reported in \cite{Chistov}, but it is still exponential-time in the number of variables, although polynomial-time in the number of polynomials and their degrees.
We note that there was a proposal some time ago for a public key cryptosystem based on the alleged hardness of inverting polynomial automorphisms \cite{Moh}, but it was subsequently attacked in \cite{Ding}.

In the present paper, we suggest to publish not all of the $\alpha(x_i)$, but only some of them, with the idea that inverting such a  ``scrap" automorphism should be harder than inverting the ``whole" $\alpha$. To the best of our knowledge, the only known way to invert a  ``scrap" automorphism is the ``brute force" trial-and-error method.

Our digital signature scheme based on ``scrap" automorphisms is described in Section \ref{automorphisms}.

Finally, we note that our digital signature scheme based on scrap matrices can be easily converted to a public key encryption scheme, see Section \ref{encryption}. We do not make any security claims about this scheme though since that would require a separate security analysis, and this is not in the scope of the present paper.

\section{Non-square matrices: scheme description}\label{description}

Let $K=\Z_q[x_1, \ldots, x_n]$ denote the algebra of polynomials in $n$ variables over the ring $\Z_q$ of integers modulo $q$, where an integer $q$ is not necessarily a prime. 

The signature scheme is as follows. 
\medskip

\noindent {\bf Public:}

\noindent -- $k \times l$ left invertible matrix $M$, with $k>l$, whose entries are sparse polynomials from the algebra $K$.\\
-- a hash function $H$ and a (deterministic) procedure for converting values of $H$ to  vectors of sparse polynomials from the algebra $K$.

\medskip

\noindent {\bf Private:}  $l \times k$ right invertible matrix $L$ over $K$, such that $LM$ is the $l \times l$ identity matrix.
\medskip

\vskip 0.2 cm

\noindent {\bf Signing} a message $m$:
\medskip

\begin{enumerate}

\item  Apply a hash function $H$ (e.g., SHA-512) to $m$. Convert $H(m)$ to a vector ${\bf U}= (P_1,\ldots, P_l)$ of $l$ (sparse) polynomials from the algebra $K$ using a deterministic public procedure.

\medskip

\item  Multiply the vector ${\bf U}$ (written as a row vector) by the (private) matrix $L$ on the right to get a vector ${\bf V}={\bf U}L= (Q_1,\ldots, Q_k)$ of $k$ polynomials from $K$.
\medskip


\item  The signature is the vector ${\bf V}$.

\end{enumerate}
\medskip

\vskip 0.2 cm

\noindent {\bf Verification.}

\medskip

\noindent {\bf 1.} The verifier computes the hash $H(m)$ and converts $H(m)$ to a vector  ${\bf U}= (P_1,\ldots, P_l)$ of $l$ (sparse) polynomials using a deterministic public procedure.
\medskip

\noindent {\bf 2.} The verifier multiplies the signature vector ${\bf V}$ by the public matrix $M$ on the right to get a vector ${\bf W}$.
\medskip

\noindent {\bf 3.} The signature is accepted if and only if ${\bf W}={\bf U}$.

\medskip

\vskip 0.2 cm

\noindent {\bf Correctness} is obvious since ${\bf W}={\bf V}M  = ({\bf U}L) M = {\bf U}(LM) = {\bf U}$.

\begin{remark}
One can use a faster, numerical, variant of the  signature verification. Namely, instead of computing a vector ${\bf W}$ of polynomials, the verifier can assign random numerical values (from $\Z_q$) to all variables $x_i$ in the vectors ${\bf U}$, ${\bf V}$ and the matrix $M$, and then do computation in $\Z_q$. To make the probability of the ``false positive" decision negligible, assigning random numerical values should be done several times.
\end{remark}

\section{Key generation}\label{generation}

In this section, we answer crucial questions on key generation in our scheme: how to generate a (left) invertible $k \times l$ matrix, how to compute its inverse, and how to generate a random sparse polynomial. We emphasize that in our scheme, all this is done offline.

\subsection{Generating a random $t$-sparse polynomial} \label{polynomial1}
We call a polynomial $t$-sparse if it has $t$ monomials. To generate a ``random" $t$-sparse polynomial, we first generate $t$ monomials as follows.
\medskip

\noindent {\bf 1.} Select the degree $d$ of a monomial, uniformly at random, from integers between 0 and $2n$, where $n$ is the number of variables.

\medskip

\noindent {\bf 2.} To select a monomial of degree $d$, we do a selection of $x_i$, uniformly at random from $\{x_1, \ldots, x_n\}$, $d$ times. Then build the monomial as a product of the selected $x_i$. 
\medskip

\noindent {\bf 3.}  Having selected $t$ monomials like that, build a $t$-sparse polynomial as a linear combination of $t$ selected  monomials with random coefficients from $\Z_q - \{0\}$.

\bigskip

\subsection{Generating a (left) invertible $k \times l$ matrix}\label{matrix}
Let $k>l$. To generate a (left) invertible $k \times l$ matrix, one can first generate an invertible square $k \times k$ matrix and then remove $k-l$ columns, selected at random.

To generate an invertible square $k \times k$ matrix, one can do the following.
\medskip

\noindent {\bf 1.} Generate an upper unitriangular $k \times k$ matrix $U$ as a product of {\it elementary matrices} $E_{ij}(u)$. A matrix $E_{ij}(u)$ has 1s on the diagonal and 0s elsewhere, except that it has a polynomial $u=u(x_1, \ldots, x_n)$ in the $(i,j)$th place, where $j > i$.

Thus, for every pair of integers $(i,j)$ with $1 \le i < j \le k$, select a random $t$-sparse polynomial $u=u_{ij}$ (as in Section \ref{polynomial1}) and make an elementary matrix $E_{ij}(u)$.

Finally, an upper unitriangular $k \times k$ matrix $U$ is computed as a product of $\frac{k^2-k}{2}$ elementary matrices selected that way.
\medskip

\noindent {\bf 2.} A lower unitriangular $k \times k$ matrix $K$ is built a similar way, except that in the elementary matrices $E_{ij}(u)$, one should have $1 \le j < i \le k$.


\medskip

\noindent {\bf 3.} An invertible $k \times k$ square matrix $S$ is now computed as a product
$U P_1 K P_2$, where $P_i$ are matrices corresponding to random permutations of columns and rows of a $k \times k$ matrix. (Entries of $P_i$ are 0s and 1s.)

\subsection{Computing the (left) inverse of a matrix}\label{inverse}

Having generated an invertible $k \times k$ square matrix $S=U P_1 K P_2$ (Section \ref{matrix}), we compute its inverse as\\ $S^{-1}=P_2^{-1}K^{-1}P_1^{-1}U^{-1}$. Computing $P_i^{-1}$ is trivial, and computing the inverse of a unitriangular square matrix $U$ or $K$ (built as in Section \ref{matrix}) is done by computing the product of inverses of the elementary matrices $E_{ij}(u)$, in the reverse order. Note that the inverse of $E_{ij}(u)$ is just $E_{ij}(-u)$.

Now suppose the (left) invertible $k \times l$ matrix $M$ was obtained from the square $k \times k$ matrix $S$ by removing $k-l$ columns $C_{i_1}, \ldots, C_{i_{k-l}}$. Then, to get a left inverse of $M$, we just remove the corresponding rows $R_{i_1}, \ldots, R_{i_{k-l}}$ from $S^{-1}$.

\subsection{Converting $H(m)$ to a vector of sparse polynomials}\label{vector}

We suggest using a hash function from the SHA-2 family, specifically SHA-512. We assume the security properties of SHA-512, including collision resistance and preimage resistance.
The following procedure may seem ``too ad hoc", but this is needed to make sure that the linear space of possible monomials obtained by this procedure has a large dimension, to prevent the forgery attack described in Section \ref{Ushakov}.

Let $B=H(m)$ be a bit string of length 512. We will convert $B$ to a vector ${\bf U}= (P_1,\ldots, P_l)$ of $l$ (4-sparse) polynomials from the algebra $K$, where each monomial has degree at most 10. We note that this process is deterministic. Of course, this is just one of the many possible conversion procedures; one of our goals here is to only have polynomials $P_i$ of low degree.

\medskip

\noindent {\bf (1)} Discard several last bits in $B$ to make its length divisible by $l$. With our suggested parameters (see Section \ref{parameters}), $l=5$, so one can  discard, say, the last 12 bits of $B$.
\medskip

\noindent {\bf (2)} Split the remaining $500$ bits in two parts: $300$ in one part and $200$ in the other. The part with $300$ bits will determine a map of the set $\{x_1, \ldots, x_{50}\}$ of variables to the set $\{x_1, \ldots, x_{63}\}$, as follows.

Each integer from 1 to 50 can be represented in the binary form using 6 bits. Thus, $300=6 \cdot 50$ bits will naturally determine a map $x_i \to x_j$ for each $i$ from 1 to 50. For example, if the first 6 bits of these 300 are 011010, this means that $x_1$ will be mapped to $x_{26}$ because 011010=26.

This map is not necessarily one-to-one (since some of the 6-bit blocks may be identical) but this is not a problem.
Let $\{x_{i_1}, \ldots, x_{i_{50}}\}$ be the new sequence of variables; some of them may be equal.
\medskip

\noindent {\bf (3)} Split the remaining $200$ bits in $l=5$ 40-bit blocks.  Then split each 40-bit block in 4 parts, 10 bits in each.  This will correspond to 4 monomials, of degree at most 10 each, as follows.  Enumerate bits in each 40-bit block by $x_{i_1}, \ldots, x_{i_{40}}$ (in this order, going left to right). Now each block of 10 bits is converted to a monomial that is a product of $x_i$ corresponding to the places in the bit string where the bit is ``1". In particular, each monomial will be of degree at most 10. For example, if the first  10-bit block is 1001100011, then the corresponding monomial will be $x_{i_1}x_{i_4}x_{i_5}x_{i_9}x_{i_{10}}$.

\medskip

\noindent {\bf (4)} Combine 4 monomials obtained at Step (3) into a 4-sparse polynomial by selecting coefficients from $\Z_q$ as follows. We are going to use the string of 12 bits that we have discarded at Step (1). Split it in 4 blocks of 3 bits. Each block will determine a coefficient at a monomial in a 4-sparse polynomial by converting a 3-bit binary number to decimal and reducing modulo 6.

\section{Suggested parameters}\label{parameters}

For the hash function $H$, we suggest SHA-512.

For the integer $q$ in $\Z_q$, we suggest $q=6$. 

For the number $n$ of variables, we suggest $n = 64$.

For the dimensions of the matrix $M$, we suggest $k=10, ~l=5$. However, due to current software limitations (see Section \ref{performance} for more details), our present implementation uses $k=5, ~l=3$.

For the number $t$ of monomials in $t$-sparse polynomials that entries of the unitriangular matrices, we suggest $t=3$.


\section{What is the hard problem here?}\label{problem}

The (computationally) hard problem that we employ in our construction is finding a left inverse of a given left invertible matrix $M$. That is, solving the matrix equation $XM=I$, where $X$ is the unknown matrix of given dimensions.

If matrices in this equation were considered over a field, then this matrix equation would translate to a system of linear equations (in the entries of the matrix $X$), and therefore would be easily solvable.

In our situation, where matrices are considered over a polynomial algebra, the problem can still be reduced to a system of linear equations, this time in the coefficients of polynomials that are entries of the matrix $X$. If the number $n$ of variables $x_i$ is not too small (we suggest $n = 64$), then the number of different monomials (and therefore the number of unknown coefficients of polynomials) is quite large (see Section \ref{linear}), so that the relevant system of linear equations becomes intractable.

\subsection{Completing a left invertible matrix to an invertible square matrix}\label{completing}

Another possible way to find a left inverse of a given left invertible matrix $M$ is to complete $M$ to an invertible square matrix by adding more columns, and then find the inverse of this square matrix (which is relatively easy since one can use determinants).

Perhaps surprisingly, it was a major open problem (Serre's problem) in algebra whether or not any left invertible matrix over a polynomial algebra can be completed to an invertible square matrix by adding more rows (or columns). This problem was answered in the affirmative independently by Quillen \cite{Quillen}  and Suslin \cite{Suslin}.

The question that matters to us though is that of the computational complexity of completing $M$ to an invertible square matrix. It was shown in \cite{Heintz} (for polynomial algebras over an infinite field) and then in \cite{Lombardi} (for polynomial algebras over an arbitrary Noetherian ring) that there is a relevant algorithm whose complexity is exponential in the square of the number of variables $x_i$. This is yet another reason why the number of variables should not be too small.

\section{Performance and signature size} \label{performance}

\noindent {\bf Disclaimer.} {\it We have run into a  problem with software for symbolic computation with multivariate polynomials, especially with matrices over multivariate polynomials. None of the symbolic computation packages either in Julia or in ANSI C (Flint) is capable of doing expand-and-simplify if the number of monomials becomes as large as several hundred. The only software that is capable of doing that is Maple, but since it is not common to use Maple for cryptographic purposes, we have decided not to use it; instead, we have reduced the values of the most impactful parameters, the dimensions of the public matrix $M$, from the recommended $10 \times 5$ to $5 \times 3$. With Flint evolving, it should be soon possible to handle $10 \times 5$ matrices over multivariate polynomials.}
\medskip

For our computer simulations, we used Apple MacBook Pro, M1 CPU (8 Cores), 16 GB RAM computer.

With the suggested parameters, signature verification takes about 0.2 sec on average, which is not bad, but the signature is rather large, about 4,200 bytes on average.

The size of the private key (the matrix $L$) is about 2,000 bytes, and so is the size of the public key (the matrix $M$).

We note that we have measured the size of a signature, as well as the size of private/public keys, as follows.
We have counted the total number of variables that occurred in relevant polynomial(s) and multiplied that number by 7, the number of bits sufficient to describe the index of any variable (except $x_{64}$). To that, we added the number of monomials times 2 (the average number of bits needed to describe a coefficient at a monomial in our construction(s)).


%
%
%

\section{Linear algebra attacks}\label{linear}

One can attempt to recover the private left inverse of the public matrix $M$ by using linear algebra, as follows. Finding the left inverse matrix of $M$ is equivalent to  solving the matrix equation $XM=I$, where $X$ is the unknown matrix of given dimensions.

If matrices in this equation were considered over a field, then this matrix equation would translate to a system of linear equations (in the entries of the matrix $X$), and therefore would be easily solvable.

In our situation, where matrices are considered over a polynomial algebra, the problem can still be reduced to a system of linear equations, this time in the coefficients of polynomials that are entries of the matrix $X$. If the number $n$ of variables $x_i$ is not too small (we suggest $n = 64$), then the number of different monomials (and therefore the number of unknown coefficients of polynomials) is very large, so that the relevant system of linear equations becomes intractable.

Note that, although monomials in the public matrix $M$ have degrees bounded by 3, in the left inverse of $M$ there can be monomials of significantly higher degree. Let us estimate the number of monomials of degree, say,  $15$ in 64 variables.

By the formula for the number of combinations with repetitions, the number is  ${78 \choose 15} \approx  2^{52}.$

Thus, the total number of potentially possible monomials in a $3 \times 6$ matrix with sparse polynomial entries is more than $2^{52} \cdot 15 > 2^{55}$, and this would be the number of unknowns in a system of linear equations. The number of equations would be at least as  large.


At least $(2^{55})^{2.3} \approx 2^{126}$ arithmetic operations are needed to solve such a system of linear equations, according to our understanding of the state-of-the-art in solving systems of linear equations.

\subsection{Forgery after collecting multiple signatures}\label{Ushakov}
A smarter linear algebra attack is targeted at forging signatures without recovering the private key and is based on collecting multiple vectors ${\bf U}=(P_1,\ldots, P_l)$ and finding a basis of the linear space generated by these vectors.

Since the transformation
${\bf U} \to {\bf V} = {\bf U} L$ is linear, after a basis of the linear space of vectors ${\bf U}$ is constructed, the forger can express each newly obtained ${\bf U}$ as a linear combination of previously collected ones, and then the signature corresponding to this newly obtained ${\bf U}$ will be the same linear combination of the relevant previously published relevant signatures ${\bf V}$.

To make this attack infeasible, the dimension of the linear space of possible vectors ${\bf U}$ should be quite large. This dimension is equal to the number of monomials that can possibly appear in polynomials $P_1,\ldots, P_l$ under the procedure of converting $H(m)$ to a vector of sparse polynomials, see Section \ref{vector}. The number of monomials of degree $k \le 64$ in 64 variables is ${64+k-1 \choose k}$, so in our situation, where the degree  can be any number between 1  and 15, the number of possible monomials is at least ${78 \choose 15} \approx 2^{52}$. This is about how many {\it linearly independent} signatures a forger has to collect. When collecting, say, 1000 linearly independent signatures a second, this would take over 50000 years.

\section{Security claims} \label{Security}

With suggested parameters, the linear algebra ``brute force" attack requires at least $2^{126}$ arithmetic  operations in $\Z_6$ to solve the relevant system of linear equations.

There could be ad hoc attacks on the public key aiming at recovering some of the entries of the private matrix, but recovering only some of the entries cannot make the probability of passing verification non-negligible because getting even just a couple of  entries (out of 15) wrong will not increase chances of  passing verification.

Forgery without getting a hold of the private key requires accumulation of at least $2^{52}$ linearly independent signatures.

As for quantum security, it appears that no known quantum algorithms are applicable in our situation since there are no abelian (semi)groups in play here that could be exploited by one of the known quantum attacks.

\section{Advantages and limitations of the matrix digital signature scheme}

\subsection{Advantages} \hfill
\medskip

\noindent {\bf 1.} Unparalleled conceptual simplicity.

\medskip

\noindent {\bf 2.} Simplicity of design. Signing a message amounts to just a matrix-vector multiplication, and so does verifying a signature.
\medskip

\noindent {\bf 3.} Bringing into play an interesting hard problem -- inverting a non-square matrix over multivariate polynomials. This problem has not been considered in the cryptographic context before, to the best of our knowledge.

\medskip

\noindent {\bf 4.} Efficiency of the signature verification (about 0.2 sec on average with parameters in the current implementation). To be fair, if we did not have to downsize dimensions of the public matrix from $10 \times 5$ to $5 \times 3$ (see the Disclaimer in Section \ref{performance}), then we project the signature verification time would be close to 1 sec.

\vskip 0.1 cm

\subsection{Limitations} \hfill

\medskip

\noindent {\bf 1.} Linearity of the signing function. To avoid linear algebra attacks, one has to use a large number of variables in the polynomial algebra, and this can affect the signature size (but not so much the verification time).
\medskip

\noindent {\bf 2.} Large signature size.

\section{Public key encryption}\label{encryption}

Our first digital signature scheme (Section \ref{description}) can be easily converted to a public key encryption scheme, as follows. We do not make any security claims about this scheme though since that would require a separate security analysis, and this is not in the scope of the present paper.
\medskip

\noindent {\bf Private:}  $k \times l$ left invertible matrix $M$, with $k>l$, whose entries are sparse polynomials from the algebra $K$.

\medskip

\noindent {\bf Public:}  $l \times k$ right invertible matrix $L$ over $K$, such that $LM$ is the $l \times l$ identity matrix.
\medskip

\vskip 0.2 cm

\noindent {\bf Encrypting} a message $m$:
\medskip

\begin{enumerate}

\item  Convert $m$ to a vector ${\bf U} = {\bf U}(m) = (P_1,\ldots, P_l)$ of $l$ (sparse) polynomials using an easily invertible public procedure. That is, presented with a vector ${\bf U}(m)$, anyone should be able to recover $m$ efficiently and without ambiguity. 

\medskip

\item  Multiply the vector ${\bf U}$ by the (public) matrix $L$ on the right to get a vector ${\bf V}= (Q_1,\ldots, Q_k)$ of $k$ polynomials from $K$. This vector ${\bf V}$ is the encryption of $m$.

\end{enumerate}

\medskip

\vskip 0.2 cm

\noindent {\bf Decryption.} Multiply the vector ${\bf V}={\bf U}L$ by the private matrix $M$ on the right to get ${\bf V}M = {\bf U}LM ={\bf U}$. Then recover $m$ from ${\bf U}$.

\section{Scrap automorphisms: scheme description}\label{automorphisms}

This scheme is similar in spirit to the scheme in Section \ref{description}. It is also similar to that scheme in some key generation details, in particular in sampling $t$-sparse polynomials. Also, converting a hash $H(m)$ to a sparse polynomial is similar, so we chose not to duplicate Section \ref{vector} here.

Again, let $K=\Z_q[x_1, \ldots, x_n]$ denote the algebra of polynomials in $n$ variables over the ring $\Z_q$ of integers modulo $q$, for an integer $q$ that is not necessarily a prime.

Let $\alpha: K \to K$ be an automorphism of the algebra $K$ given by $n$ polynomials $\alpha(x_i)$.
The signature scheme is as follows.
\medskip

\noindent {\bf Public:} $k$ polynomials $y_{i_1}=\alpha(x_{i_1}), \ldots, y_{i_k}=\alpha(x_{i_k})$, where $k<n$.
\medskip

\noindent {\bf Private:} The inverse $\alpha^{-1}$ of the automorphism $\alpha$, given by $n$ polynomials $z_i = \alpha^{-1}(x_i)$.
\medskip

\vskip 0.2 cm

\noindent {\bf Signing} a message $m$:
\medskip

\noindent {\bf 1.} Apply the hash function $H$ (e.g., SHA-512) to $m$. Convert $H(m)$ to a sparse polynomial $Q$ in $y_{i_1}, \ldots, y_{i_k}$ using a deterministic public procedure (see e.g. Section \ref{vector}).
\medskip

\noindent {\bf 2.} Apply the automorphism $\alpha^{-1}$ to the polynomial $Q$; the result is
$Q(x_{i_1}, \ldots, x_{i_k})$. Denote the latter polynomial by $S$.
\medskip

\noindent {\bf 3.} The signature is the polynomial $S=\alpha^{-1}(Q)$.
\bigskip

\noindent {\bf Verification:}

\medskip

\noindent {\bf 1.} The verifier computes $H(m)$ and converts $H(m)$ to $Q=Q(y_{i_1}, \ldots, y_{i_k})$ using a deterministic public procedure.

\medskip

\noindent {\bf 2.} The verifier applies the public automorphism $\alpha$ to the polynomial
$S=Q(x_{i_1}, \ldots, x_{i_k})$.
\medskip

\noindent {\bf 3.} The signature is accepted if and only if $\alpha(S)$ coincides with
$Q = Q(y_{i_1}, \ldots, y_{i_k})$.
\medskip

\vskip 0.2 cm

\noindent {\bf Correctness} is obvious since $\alpha(S) = \alpha(\alpha^{-1}(Q)) = \alpha(Q(x_{i_1}, \ldots, x_{i_k})) = Q(y_{i_1}, \ldots, y_{i_k})$.

%

\subsection{Generating an automorphism $\alpha$}\label{automorphism}

An automorphism $\alpha$ is generated offline, as follows. Recall that a polynomial is {\it $t$-sparse} if it only has $t$ monomials.
\medskip

\noindent {\bf (1)}  Choose an index $k$ at random between 1 and $n$, where $n$ is the number of variables in our polynomial algebra.
\medskip

\noindent {\bf (2)} Take $x_k$ to $x_k + h(x_1,\ldots, x_n)$, where $h(x_1,\ldots, x_n)$ is a random $t$-sparse polynomial not depending on $x_k$. Fix all other variables.
\medskip

\noindent {\bf (3)} Apply a random permutation to the set $x_1,\ldots, x_n$ of variables.
\medskip

\noindent {\bf (4)} Repeat steps (1) through (3) $l$ times (we suggest $l=n$).
\medskip

Keep record of all ``elementary" automorphisms (Steps (2) and (3)) used in the process of building $\alpha$.
\medskip

\subsection{Computing the automorphism $\alpha^{-1}$}\label{inverse}

Having generated the automorphism $\alpha$ (see the previous subsection), we now have to compute
$\alpha^{-1}$, the private key of the signatory.

To do that, we trace back the ``elementary" automorphisms (see Section \ref{automorphism}) and invert them. More specifically,
if $\alpha$ is a composition $\alpha=\varepsilon_1 \cdots \varepsilon_r$ (with $\varepsilon_1$  applied first), then $\alpha^{-1} = \varepsilon_r^{-1} \cdots \varepsilon_1^{-1}$.

To invert a permutation is straightforward, and if $\varepsilon_j$ is as in Step (2) in the previous Section \ref{automorphism}, then $\varepsilon_j^{-1}$ takes $x_k$ to $x_k - h(x_1,\ldots, x_n)$ and fixes all other variables.

\subsection{Sampling a $t$-sparse polynomial}\label{polynomial}

The number $t$ of monomials is one of the parameters of the scheme, see Section \ref{parameters2}.
Sampling a $t$-sparse polynomial can be done the same way as in Section \ref{polynomial}.
\medskip

\subsection{Suggested parameters}\label{parameters2}

Again, for the integer $q$ in $\Z_q$, we suggest $q=6$.
\medskip

\noindent For the number $n$ of variables, we suggest $n = 32$.
\smallskip

\noindent For the number $k$ of published polynomials $y_{i}=\alpha(x_{i})$, we suggest $k=16$.
\smallskip

\noindent For the number $t$ of monomials in $t$-sparse polynomials, we suggest $t=3$.
\smallskip

\noindent For the degree bound $b$ of monomials in $t$-sparse polynomials, we suggest $b=3$.
\smallskip

\noindent For the number $s$ of elementary automorphisms that compose to an automorphism $\alpha$, we suggest $s=16$.

\subsection{Estimating the size of the key space with suggested parameters}\label{keyspace2}

As we have mentioned in the Introduction, the only known way to invert a ``scrap" automorphism is the ``brute force" trial-and-error method. It is therefore natural to ask for a complexity estimate of such a method. In other words, what is the size of the key space that a brute force attacker would have to exhaust?

First, let us estimate the number of monomials of degree $d\le 3$ in 32 variables.
By the formula for the number of combinations with repetitions, this number is ${34 \choose 3} + {33 \choose 2} + {31 \choose 1} = 6,543$.

Thus, the number of $t$-sparse polynomials with $t=3$ is at least ${6,543 \choose 3} \cdot 5^3  > 2^{42}$. (The factor $5^3$ appears because our polynomial is a linear combination of 3 monomials with coefficients from $\Z_6 - \{0\}$.) Then the number of ``elementary" automorphisms (as in Section \ref{automorphism})
is at least $2^{42} \cdot 32 = 2^{47}$. The number of compositions of $s=16$ elementary  automorphisms is therefore at least $(2^{42})^{16}= 2^{672}$.

\bigskip

\noindent {\bf Acknowledgement.} We are grateful to Sasha Ushakov for useful discussions, in particular for pointing out the forgery attack in Section \ref{Ushakov}.

\end{document}